\documentclass[11pt]{article}
\usepackage{asp2006}
\usepackage{epsf}
\usepackage{graphics}
\def\edcomment#1{\iffalse\marginpar{\raggedright\sl#1\/}\else\relax\fi}
\reversemarginpar

\begin{document}
\title{Circumnuclear Regions of Star Formation in Early Type Galaxies}
\author{\'Angeles I. D\'\i az\altaffilmark{1}, Elena Terlevich\altaffilmark{2}, Guillermo F. H\"agele\altaffilmark{1}, Marcelo Castellanos\altaffilmark{1}}
\altaffiltext{1}{Departamento de F\'\i sica Te\'orica, Universidad Aut\'onoma de Madrid, Spain}
\altaffiltext{2}{Instituto Nacional de Astrof\'\i sica, \'Optica y Electr\'onica, Puebla, M\'exico}

\begin{abstract}
Circumnuclear star forming regions, also called hotspots, are often found in the inner regions of some spiral galaxies where intense processes of star formation are taking place. In the UV, massive stars dominate the observed circumnuclear emission even in the presence of an active nucleus, contributing between 30 and 50 \% to the H$\beta$ total emission of the nuclear zone.
Spectrophotometric data of moderate resolution ( 3000 $<$ R $<$ 11000) are presented from which the physical properties of the ionized gas: electron density, oxygen abundances, ionization structure etc. have been derived. 
\end{abstract}

\vspace{-0.5cm}
\section{Introduction}
The inner ($\sim$ 1Kpc) zones of some spiral galaxies show high star formation rates frequently arranged in a ring or pseudo-ring pattern around their nuclei. 
In general, Circumnuclear Star Forming Regions (CNSFR) -- also referred to as ``hotspots'' -- and luminous and large disk HII regions are very much alike, but the first look more compact and show higher peak surface brightness  (Kennicutt et al. 1989).  In many cases they contribute substantially to the emission of the entire nuclear region. 
Their large H$\alpha$ luminosities, typically higher than 10$^{39}$ erg s$^{-1}$, point to relatively massive star clusters as their ionisation source. 
These regions then constitute excellent places to study how star formation proceeds in high metallicity, high density circumnuclear environments.

The importance of an accurate determination of the abundances of high metallicity HII regions cannot be overestimated since they constitute most of the HII regions in early spiral galaxies (Sa to Sbc) and the inner regions of most late  type ones (Sc to Sd) without  which our description of the metallicity distribution in galaxies cannot be  complete.  The question of how high is the highest oxygen abundance in the gaseous phase of galaxies is still standing and extrapolation of known radial abundance gradients would point to CNSFR as the most probable sites for these high metallicities. 
Accurate measurements of elemental abundances of high metallicity regions are therefore crucial to obtain reliable calibrations of empirical abundance estimators, widely used but poorly constraint, whose choice can severely bias results obtained for quantities of the highest relevance for the study of galactic evolution like the luminosity-metallicity (L-Z) relation for galaxies. CNSFR are also ideal cases for studying the behaviour of abundance estimators in the high metallicity regime. 

\section{Observations and measurements}
We have obtained moderate resolution observations of 12 CNSFR in three ``hot spot" galaxies: NGC~2903, NGC~3351 and NGC~3504. The three of them are early barred spirals and show  high star formation rate in their nuclear regions. They are quoted in the literature as among the spirals with the highest overall oxygen abundances (
P\'erez-Olea 1996). Figure 1 shows an archive HST telescope image of the circumuclear region of one of  the observed galaxies, NGC~ 3351, taken with the WFPC2 camera through the F606W filter. 

The observations were obtained with the ISIS double spectrograph mounted on the 4.2m WHT at the Roque de los Muchachos Observatory and provided spectrophotometry in the  spectral ranges  $\lambda$3650
- $\lambda$7000 {\AA} in the blue and  ($\lambda$8850 -$\lambda$9650) in the near IR, with spectral resolutions of {$\sim$}2.0 {\AA} and 1.5 {\AA} FWHM respectively. Different slit positions were used to observe several distinct HII regions as exemplified in Figure 1.  The data were reduced using the IRAF (Image Reduction and Analysis Facility) package following standard procedures. The blue and red spectra corresponding to one of the observed regions are also shown. 
The presence of a conspicuous underlying stellar population
in most observed regions, 
more evident in the blue spectra, complicates the measurements of emission lines (see blue spectrum in Figure 1). A two-component (emission and absorption) gaussian fit was performed in order to correct the Balmer lines for underlying absorption. 

\begin{figure}[!ht]
\plotone{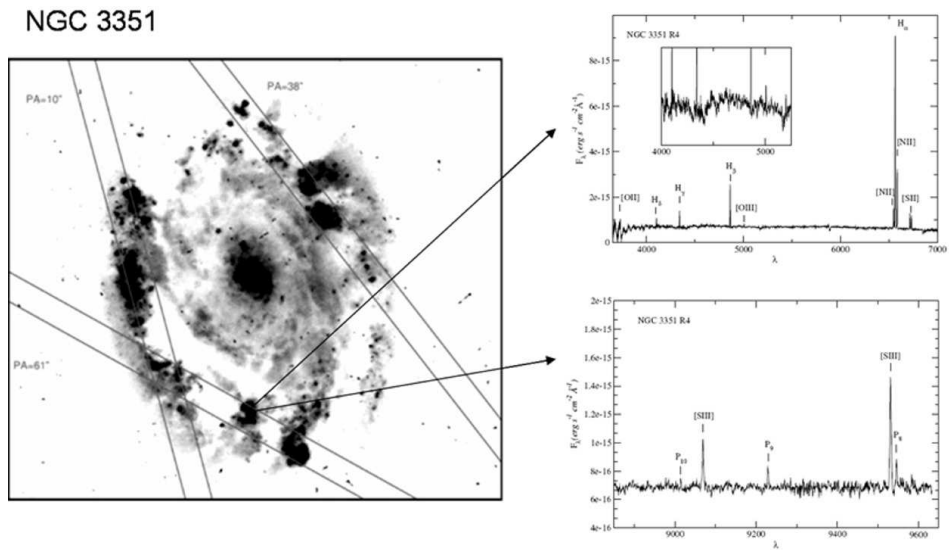}
\caption{Observed CNSFR in NGC~3351. The different slit positions are superimposed on images taken from the HST archive and obtained with the WFPC2 camera through the F606W filter. The position angles of every slit position are indicated. The blue and red spectra corresponding to one of the observed regions are also shown. }
\end{figure}

\section{Results and discussion}
Electron densities for each observed region have been derived from the  [SII] $\lambda\lambda$ 6717, 6731 \AA\ line ratio, following standard methods. They were found to be, in all cases, 
$\le$ 600 cm$^{-3}$, higher than those usually derived in disk HII regions, but still below the critical value for collisional de-excitation. The low excitation of the regions, as evidenced by the weakness of the [OIII] $\lambda$ 5007 \AA\ line (see the blue spectrum in Figure 1, precludes the detection and measurement of the auroral [OIII] $\lambda$ 4363 \AA\ necessary for the derivation of the electron temperature. We have therefore used a semi-empirical procedure for the derivation of abundances. Firstly we have produced a calibration of the [SIII] temperature as a function of the abundance parameter SO$_{23}$ defined as: 
\[
SO_{23} = \frac{I([OII]\lambda 3727,29)+I([OIII]\lambda 4959,5007)}{I([SII]\lambda 6716,31)+I([SIII]\lambda 9069,9532)}
\]
This parameter is similar to the S$_3$O$_3$ proposed by Stasi\'nska (2006) but is, at first order, independent from geometrical (ionization parameter) effects. 
To perform this calibration we have compiled all the data so far available with sulfur emission line detections of both the auroral and nebular lines at $\lambda$ 6312 \AA\ and $\lambda\lambda$ 9069,9532 \AA\ respectively (see D\'\i az et al. 2007).
The calibration is shown in Figure 2 together with the quadratic fit to the high metallicity HII region data:
\[
t_e([SIII]) = 0.596 - 0.283 log SO_{23} + 0.199 (log SO_{23})^2
\]
For one of the observed regions, R1+R2, the [SIII] $\lambda$ 6312 \AA\  line has been detected and measured yielding a value T$_e$([SIII])= 8400$^{+ 4650}_{-1250}$K, represented as a solid black circle in Figure 2 (left).
We have used this calibration to derive t$_{e}$([SIII]) for our observed CNSFR. In all cases the values of log$O_{23}$ are inside the range used in performing the calibration, thus requiring no extrapolation of the fit.These temperatures, in turn, have been used to derive the S$^+$/H$^+$ and S$^{++}$/H$^+$ ionic ratios.
Once the sulphur ionic abundances have been derived, we have estimated the corresponding oxygen abundances, assuming that sulphur and oxygen temperatures follow the relation given by Garnett (1992) and t$_e$ ([OII]) $\simeq$ t$_e$([SIII]).  Finally, we have derived the N$^{+}$/O$^{+}$ ratio assuming that t$_e$([OII]) $\simeq$ t$_e$([NII]) $\simeq$ t$_e$([SIII]).

 \begin{figure}[!h]
 \plottwo{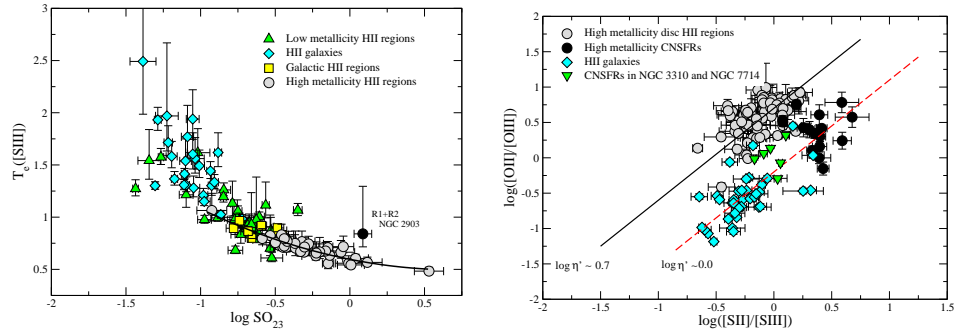}{eta-prima-plot.eps}
 \caption{{\itshape Left:\/}Empirical calibration of the [SIII] electron temperature as a function of the abundance parameter SO$_{23}$. The solid line represents a quadratic fit to the high metallicity HII region data.
 {\itshape Right:\/} The [OII]/[OIII] vs [SII]/[SIII] logarithmic ratios for different ionised regions. References for the data in the two figures can be found in D\'\i az et al. (2007)}
 \end{figure}

The observed CNSFR being of high metallicity show however marked differences with respect to high metallicity disk HII regions. Even though their derived oxygen and sulphur abundances are similar, they show values of the O$_{23}$ and the N2 parameters whose distributions are shifted to lower and higher values respectively with respect to the high metallicity disk sample. Hence, if pure empirical methods were used to estimate the oxygen abundances for these regions, higher values would in principle be obtained. This would seem to be in agreement with the fact that CNSFR, when compared to the disk high metallicity regions, show the highest [NII]/[OII] ratios. 

 CNSFR also show lower ionization parameters than their disk counterparts, as derived from the [SII]/[SIII]  ratio. Their ionization structure also seems to be different with CNSFR showing radiation field properties more similar to HII galaxies than to disk high metallicity HII regions. This can be seen in Figure 2 (right), a diagram of the emission line ratios [OII]/[OIII] vs [SII]/[SIII], that works as a diagnostics for the nature and temperature of the radiation field. In this plot, diagonal lines of slope unity show the locus of ionized regions with constant ionization temperature. CNSFRs are seen to segregate from disk HII regions. The former cluster around a value of T$_{ion}\sim$ 40,000 K, while the latter cluster around T$_{ion}\sim$ 35,000 K. Also shown are the data corresponding to HII galaxies. Indeed, CNSFRs seem to share more the locus of HII galaxies than that of disk HII regions.

One possible concern about these CNSFR is that, given their proximity to the galactic nuclei, they could be affected by hard radiation coming from a low luminosity AGN.
 Alternatively, the spectra of  these regions harbouring massive clusters of young stars might be affected by the presence of shocked gas. Diagnostic diagramas of the kind presented by Baldwin, Phillips, \& Terlevich (1981) can be used to investigate the possible contribution by either a hidden AGN or the presence of shocks to the emission line spectra of the observed CNSFR. When plotted on one of these diagnostic diagramas, log ([NII]/H$\alpha$) vs log ([OIII])/H$\beta$, some of our CNSFR are found close to the transition zone between HII region and LINER spectra but only one region in NGC~3504 may show a hint of a slight contamination by shocks. 

\acknowledgements{We acknowledge finantial support from DGICYT grant AYA-2004-02860-C03. AID thanks the hospitality of the Institute of Astronomy, Cambridge, during her sabbatical finnanced by the Spanish MEC through 
grant PR2006-0049.

\end{document}